\documentclass[10pt]{iopart}
\usepackage{graphicx}
\usepackage{color}

\expandafter\let\csname equation*\endcsname\relax

\expandafter\let\csname endequation*\endcsname\relax

\usepackage[libertine]{newtxmath}
\usepackage{slashed}
\usepackage{amsfonts}

\usepackage[normalem]{ulem}

\usepackage{microtype}\usepackage[colorlinks,linkcolor=blue,anchorcolor=blue,citecolor=blue,urlcolor=blue,breaklinks=true]{hyperref}

\begin{document}

\title{Constraints on the dark sector from electroweak precision observables}

\author{B~M~Loizos, X~G~Wang, A~W~Thomas, M~J~White and A~G~Williams}

\address{CSSM and ARC Centre of Excellence for Dark Matter Particle Physics, Department of Physics, The University of Adelaide, Adelaide, Australia}
\ead{\mailto{bill.loizos@adelaide.edu.au}, \mailto{xuan-gong.wang@adelaide.edu.au}, \mailto{anthony.thomas@adelaide.edu.au}, \mailto{martin.white@adelaide.edu.au} and \mailto{anthony.williams@adelaide.edu.au}}
\vspace{10pt}
\begin{indented}
\item[]March 2024
\end{indented}

\begin{abstract}
We revisit the Standard Model fit to electroweak precision observables using the latest data and the Particle Data Group value of the mass of the W boson. This analysis is repeated for the value reported by CDF. The constraints on the parameter space for dark photons arising from these electroweak precision observables are then evaluated for both values of the W boson mass. We also extend previous work by placing the first electroweak precision observable constraints on the coupling of dark photons to the fermionic dark matter sector.
\end{abstract}

%
%
%
%
%

\section{Introduction}
When developing extensions of the Standard Model (SM) to include potential particle candidates for dark matter (DM), the possibility that a portal exists that bridges the
DM sector with SM particles is an enticing one. One interesting example of such a portal is the so-called dark photon. This couples to weak hypercharge before electroweak symmetry breaking and so mixes with both the photon and the $Z$ boson and hence couples to other SM particles. 
The dark photon may also couple to
some or all potential dark matter particles in the dark sector. Here we consider the case of dark Dirac fermions.
The model is described in the following section. A number of recent reviews have explored the theoretical implications of the presence of a dark photon along with the existing and anticipated experimental constraints on its allowed parameter space~\cite{Fabbrichesi:2020wbt, Filippi:2020kii, Graham:2021ggy}.

Numerous experimental searches for the dark photon have been undertaken~\cite{Merkel:2014avp, LHCb:2019vmc, Banerjee:2019pds, BaBar:2017tiz, CMS:2019buh}. While there is no direct evidence so far, a recent study of world data on deep inelastic scattering (DIS) did report indirect evidence for its existence in the few GeV 
 range~\cite{Hunt-Smith:2023sdz}. The NA64~\cite{Banerjee:2019pds} and BaBar~\cite{BaBar:2017tiz} experiments have placed strong constraints on the kinetic mixing parameter, $\epsilon \le 10^{-3}$, for a dark photon mass up to $8\ {\rm GeV}$, albeit with small gaps around the $J/\Psi$ and its excitations. 
 The CMS Collaboration~\cite{CMS:2019buh} has derived similarly competitive limits in the heavy mass region. The sensitivity of these limits has recently been re-examined in light of the potential coupling of the dark photon to dark matter~\cite{Abdullahi:2023tyk}.
 There are also several planned experiments~\cite{APEX:2011dww,Battaglieri:2014hga, Beranek:2013yqa} aiming to explore parts of the remaining allowed parameter space.
 
 Theoretical investigations have also placed decay-agnostic constraints on the dark photon parameters by exploring the consequences for many physical processes and then using relevant experiments to constrain parameters. The physics processes considered  
 include measurements of the $g-2$ of the muon~\cite{Pospelov:2008zw, Davoudiasl:2012qa}, electroweak precision observables (EWPO)~\cite{Hook:2010tw, Curtin:2014cca}, 
 $e^- p$ DIS~\cite{Kribs:2020vyk, Thomas:2021lub, Yan:2022npz}, parity violating electron scattering~\cite{Thomas:2022qhj, Thomas:2022gib}, partial wave unitarity~\cite{Hosseini:2022urq},
rare kaon and $B$-meson decays~\cite{Davoudiasl:2012ag, Wang:2023css}, and high-luminosity LHC projections~\cite{McCullough:2022hzr}.

The dark photon framework has also been applied to ascertain whether it could explain the $W$ boson mass anomaly reported by  the CDF Collaboration~\cite{CDF:2022hxs}. 
The favoured regions of dark photon parameter space were derived by fitting the CDF $m_W$~\cite{Thomas:2022gib, Zhang:2022nnh, Zeng:2022lkk, Cheng:2022aau}. In this work, we will investigate how the CDF $m_W$ affects the previous EWPO exclusion limits of dark photon parameters~\cite{Curtin:2014cca}, by a global fit to $m_W$ and the other 16 electroweak precision observables.
The impact of the CDF $m_W$ measurement on the fit of electroweak data in the Standard Model and beyond has been investigated recently in Refs.~\cite{Lu:2022bgw, deBlas:2022hdk}, including studies of new physics models with oblique corrections, the two-Higgs doublet model and dimension six Standard Model effective field theory (SMEFT).

In addition to the correction a dark photon may generate for EWPO, it is also possible that it could serve as a portal to a new DM sector that would otherwise not interact with SM particles~\cite{Boehm:2003hm, Hambye:2019dwd}. 
There are many well-motivated dark matter candidates for this sector, with masses ranging from ultra-light~\cite{Hu:2000ke, Veltmaat:2016rxo} to super-heavy~\cite{Luongo:2018lgy, Carney:2022gse}.
For a DM sector consisting of dark photons coupling to light DM particles, sub-GeV Dirac fermions have been ruled out by the Planck data~\cite{Planck:2015fie}, while cosmological constraints have been placed on the variable $y = \epsilon^2 \alpha_D (m_{\chi}/m_{A'})^4$ for scalar, pseudo-Dirac, and asymmetric DM scenarios~\cite{Izaguirre:2015yja}, where $m_{A'}$ and $m_{\chi}$ are the masses of dark photon and dark matter particles, respectively, and $\alpha_D = g_\chi^2/4\pi$ and $g_\chi$
is the coupling strength of the dark photon to the dark sector particles. 
In typical analyses, $\alpha_D$ is flexible and can vary up to the perturbativity bound. However, it is also important to attempt to place constraints on $g_{\chi}$ directly, as we do here.

In this work, we revisit the electroweak constraints on the dark photon parameters. 
We first examine the implication of the CDF $m_W$ measurement on the EWPO constraints on the dark photon kinetic mixing parameter $\epsilon$. Then we set exclusion limits on the dark photon couplings to dark matter particles in the case of Dirac fermions with $m_{\chi}$ up to $m_Z/2$.

In Sec.~\ref{sec:darkphoton}, we begin by briefly reviewing the dark photon formalism, including the coupling to dark fermions. In Sec.~\ref{sec:EWPO} we consider the global fit to electroweak precision observables within the SM. In Sec.~\ref{sec:constraints} we present the exclusion limits on dark photons and dark fermions, and finally we summarise our conclusions in Sec.~\ref{sec:conclusion}.

\section{Dark photon formalism} 
\label{sec:darkphoton}
The dark photon is usually introduced as an extra $U(1)$ gauge boson~\cite{Fayet:1980ad, Fayet:1980rr, Holdom:1985ag}, 
interacting with SM particles through kinetic mixing with hypercharge~\cite{Okun:1982xi}
\begin{equation}
\label{eq:L}
 {\cal L}  \supset  
- \frac{1}{4} F'_{\mu\nu} F'^{\mu\nu} + \frac{1}{2} m^2_{A'} A'_{\mu} A'^{\mu} 
+ \frac{\epsilon}{2 \cos\theta_W} F'_{\mu\nu} B^{\mu\nu} + g_{\chi} \bar{\chi} \gamma^{\mu} \chi A'_{\mu} \, ,
\end{equation}
where $\theta_W$ is the weak mixing angle, $F'_{\mu\nu}$ is the dark photon strength tensor and $\epsilon$ is the mixing parameter.
We use $A'$ and $\bar{Z}$ to denote the unmixed versions of the dark photon and the SM neutral weak boson, respectively. Note here that we also introduced the minimal coupling of the $A'$ to dark fermions $\chi$. 

After diagonalising the mixing term through the following field redefinitions,
\begin{eqnarray}
B_{\mu} &\rightarrow& B_{\mu} + \frac{\epsilon}{\cos\theta_W} A'_{\mu}\, , \nonumber\\
A'_{\mu} &\rightarrow& \frac{1}{\sqrt{1 - \epsilon^2/\cos^2\theta_W}} A'_{\mu} \, ,
\end{eqnarray}
the dark photon and $\bar{Z}$ mass-squared matrix becomes,
\begin{eqnarray}
M^2= m_{\bar{Z}}^2\left[\begin{array}{cc}
1 & -\epsilon_{\mathrm{W}} \\
-\epsilon_{\mathrm{W}} & \epsilon_{\mathrm{W}}^2+\rho^2
\end{array}\right]\, ,
\end{eqnarray}
where
\begin{eqnarray}
\label{eq:epsW-rho}
\epsilon_W &=& \frac{\epsilon \tan \theta_W}{\sqrt{1 - \epsilon^2/\cos^2\theta_W}} ,\nonumber\\
\rho &=& \frac{m_{A'}/m_{\bar{Z}}}{\sqrt{1 - \epsilon^2/\cos^2\theta_W}} \, .
\end{eqnarray}
By diagonalising the mass-squared matrix, one can define the physical $Z$ and $A_D$,
\begin{eqnarray}
\label{eq:Z-AD}
Z_{\mu} &=& \cos\alpha \bar{Z}_{\mu} + \sin\alpha A'_{\mu} \, , \nonumber\\
A_{D\mu} &=& - \sin\alpha \bar{Z}_{\mu} + \cos\alpha A'_{\mu}\, ,
\end{eqnarray}
where $\alpha$ is the $\bar{Z}-A'$ mixing angle,
\begin{equation}
\tan \alpha = \frac{1}{2\epsilon_W} \Big[ 1 - \epsilon^2_W - \rho^2 
- {\rm sign}(1-\rho^2) \sqrt{4\epsilon_W^2 + ( 1 - \epsilon_W^2 - \rho^2)^2} \Big] \, .
\end{equation}
The masses of these physical states are~\cite{Kribs:2020vyk}
\begin{equation}
\label{eq:m_Z_AD}
m^2_{Z, A_D} = \frac{m_{\bar{Z}}^2}{2} [ 1 + \epsilon_W^2 + \rho^2 \pm {\rm sign}(1-\rho^2) \sqrt{(1 + \epsilon_W^2 + \rho^2)^2 - 4 \rho^2} ] \, .
\end{equation}
Due to kinetic mixing, the SM weak couplings of the $Z$ boson to both leptons and quarks will be modified, and the dark photon will also couple to SM particles. In our framework, both $Z$ and $A_D$ will couple to dark fermions.
The interacting Lagrangian becomes
\begin{eqnarray}
\label{eq:LV}
{\cal L}_V &=& 
- e Q_f \bar{f} \gamma^{\mu} f A_{\mu} 
- \frac{e}{\sin2\theta_W} \bar{f}\gamma^{\mu} (C^v_Z - C^a_Z \gamma_5) f Z_{\mu} \nonumber\\
&& - \frac{e}{ \sin2\theta_W} \bar{f} \gamma^{\mu} (C^v_{A_D} - C^a_{A_D} \gamma_5) f A_{D\mu}\nonumber\\
&& + \frac{g_{\chi}}{\sqrt{1-\epsilon^2/\cos^2\theta_W}} (\bar{\chi} \gamma^{\mu} \chi) (Z_{\mu}\sin\alpha + A_{D\mu}\cos\alpha)\, ,\nonumber\\
\end{eqnarray}
where $Q_f$ is the electric charge of the SM fermion $f$.

There are three independent parameters, $m_{\bar Z}$, $m_{A'}$ and $\epsilon$. Alternatively, we can choose the physical mass $m_{A_D}$ as a parameter instead of $m_{A'}$.
From Eq.~(\ref{eq:m_Z_AD}), we rewrite $\rho^2$ in terms of $m_{\bar Z}$, $m_{A_D}$ and $\epsilon$ as
\begin{eqnarray}
\rho^2 = 
\dfrac{m^2_{\bar Z}}{m^2_{A_D} - m^2_{\bar Z}} \Big[ \dfrac{m^4_{A_D}}{m^4_{\bar Z}} - \dfrac{m^2_{A_D}}{m^2_{\bar Z}} \left( 1 + \epsilon^2_W \right) \Big]\, .
\end{eqnarray}

All the physical couplings in Eq.~(\ref{eq:LV}) depend on three parameters: $m_{\bar Z}$, $m_{A_D}$, and $\epsilon$.
In other analyses~\cite{Thomas:2021lub, Thomas:2022gib, Hunt-Smith:2023sdz, Thomas:2022qhj, Wang:2023css} where the physical $m_Z$ is fixed at its experimental value, only two parameters are independent, with $m_{A_D}$ and $\epsilon$ as the usual choice.

\subsection{Couplings to SM fermions}
\label{sec:C_SM}
The lowest order SM couplings of the $Z$ boson to leptons and quarks, $C_{\bar{Z}}^v = \{g^{\nu}_V, g^e_V, g^u_V, g^d_V\}$ and $C_{\bar{Z}}^a = \{ g^{\nu}_A, g^e_A, g^u_A, g^d_A\}$, will be shifted to~\cite{Kribs:2020vyk,Thomas:2022qhj,Wang:2023css}~\footnote{In our notation, the couplings $C_{\bar Z}$, $C_{Z}$ and $C_{A_D}$ are different from those in Ref. [18] by a factor of $\sin2\theta_W$.}
\begin{eqnarray}
\label{eq:C_Z}
C_Z^v &=& (\cos\alpha - \epsilon_W \sin\alpha) C_{\bar{Z}}^v + 2 \epsilon_W \sin\alpha \cos^2 \theta_W C_{\gamma}^v ,\nonumber\\
C_Z^a &=& (\cos\alpha - \epsilon_W \sin\alpha) C_{\bar{Z}}^a ,
\end{eqnarray}
where 
$C_{\gamma}^v = \{ C^{\nu}_{\gamma}, C^e_{\gamma}, C^u_{\gamma}, C^d_{\gamma}\} = \{ 0, -1, 2/3, - 1/3 \}$. 

Likewise, the couplings of the physical dark photon $A_D$ to SM fermions are given by
\begin{eqnarray}
\label{eq:C_AD}
C_{A_D}^v &=& - (\sin\alpha + \epsilon_W \cos\alpha) C_{\bar{Z}}^v + 2 \epsilon_W \cos\alpha \cos^2 \theta_W C_{\gamma}^v ,\nonumber\\
C_{A_D}^a &=& - (\sin\alpha + \epsilon_W \cos\alpha) C_{\bar{Z}}^a 
\, .
\end{eqnarray}
%

\subsection{Couplings to dark fermions}
Both the dark photon $A_D$ and the physical $Z$ boson will couple to dark matter particles. From Eq.~(\ref{eq:LV}), we can define the effective couplings
\begin{eqnarray}
\label{eq:C-chi-chi}
C^v_{A_D, \chi\bar{\chi}} &=& \frac{g_{\chi} \cos\alpha}{\sqrt{1- \epsilon^2/\cos^2\theta_W}}\, ,\nonumber\\
C^v_{Z,\chi\bar{\chi}} &=& \frac{g_{\chi} \sin\alpha}{\sqrt{1- \epsilon^2/\cos^2\theta_W}}\, .
\end{eqnarray}
Note that in the minimal $U(1)_X$ model, where the dark photon $A'$ only kinetically mixes with the physical photon, the $Z$ boson will not couple to dark matter particles.

\section{Electroweak observables}
\label{sec:EWPO}

\subsection{Effective couplings}

The $Z$-pole observables can be expressed in terms of the effective couplings~\cite{Hagiwara:1994pw, Cho:1999km}, which were derived in the $\overline{{\rm MS}}$ renormalization scheme.
The effective vector- and axial-vector couplings, $v_f$ and $a_f$,
of the Z boson to leptons at
the Z-pole $i \bar f \gamma^\mu (v_f - a_f\gamma_5)f\,Z_\mu$, can be parameterised by~\cite{Cho:1999km, Cho:2011rk},
\begin{eqnarray}
\label{eq:v-a-fermion}
v_{\nu} &=& g^{\nu}_{V} + \left( 0.00199 + 0.45250 \Delta {\bar g}^2_Z + 0.00469 \Delta {\bar s}^2  \right)\, ,\nonumber\\
a_{\nu} &=& g^{\nu}_{A} + \left( 0.00199 + 0.45250 \Delta {\bar g}^2_Z + 0.00469 \Delta {\bar s}^2  \right)\, ,\nonumber\\
v_e &=&  g^e_V + \left( 0.00033 - 0.03426 \Delta {\bar g}^2_Z + 2.01197 \Delta {\bar s}^2 \right)\, ,\nonumber\\
a_e &=&  g^e_A +  \left( - 0.00127 - 0.45250 \Delta {\bar g}^2_Z - 0.00371 \Delta {\bar s}^2 \right)\, ,\nonumber\\
v_u &=&  g^u_V +  \left( 0.00041 + 0.17367 \Delta {\bar g}^2_Z - 1.33977 \Delta {\bar s}^2 \right)\, ,\nonumber\\
a_u &=&  g^u_A +  \left( 0.00145 + 0.45251 \Delta {\bar g}^2_Z + 0.00391 \Delta {\bar s}^2 \right) \, ,\nonumber\\
v_d &=&  g^d_V +  \left( - 0.00118 - 0.31308 \Delta {\bar g}^2_Z + 0.66756 \Delta {\bar s}^2 \right)\, ,\nonumber\\
a_d &=&  g^d_A +  \left( - 0.00168 - 0.45250 \Delta {\bar g}^2_Z - 0.00424 \Delta {\bar s}^2 \right)\, ,
\end{eqnarray}
where 
\begin{eqnarray}
\Delta {\bar s}^2 &=& 0.00360 \Delta S_Z - 0.00241 \Delta T_Z + 0.00011 x_{\alpha}\, ,\nonumber\\
\Delta {\bar g}^2_Z &=& 0.00412 \Delta T_Z\, ,
\end{eqnarray}
with
\begin{eqnarray}
\Delta S_Z &=&  0.2217 x_h - 0.1188 x_h^2 + 0.0320 x_h^3 - 0.0014 x_t + 0.0005 x_s\, ,\nonumber\\
\Delta T_Z &=& - 0.0995 x_h - 0.2858 x_h^2 + 0.1175 x_h^3 + 0.0367 x_t + 0.00026 x_t^2 \nonumber\\ 
&& - 0.0017 x_h x_t - 0.0033 x_s - 0.0001 x_t x_s\, ,
\end{eqnarray}
where
\begin{eqnarray}
x_h &=& \dfrac{\ln(m_h/100\ {\rm GeV})}{\ln 10}\, ,\ x_t = \frac{m_t - 172}{3}\, ,\nonumber\\
\ x_s &=& \frac{\alpha_s(m_Z) - 0.118}{0.003}\, ,\ x_{\alpha} = \frac{\Delta\alpha^{(5)}_{\rm had} - 0.0277}{0.0003}\, .
\end{eqnarray}
For the b quark, the $Zb{\bar b}$ couplings have non-trivial $m_t$ and $m_h$ dependence due to large vertex corrections~\cite{Cho:2011rk},
\begin{eqnarray}
\label{eq:v-a-bquark}
v_b &=&  g^b_V +  \left( 0.00208 - 0.31308 \Delta {\bar g}^2_Z + 0.66756 \Delta {\bar s}^2 \right) + \Delta g^b_{V}\, ,\nonumber\\
a_b &=&  g^b_A +  \left( 0.00142 - 0.45250 \Delta {\bar g}^2_Z - 0.00424 \Delta {\bar s}^2 \right) + \Delta g^b_{A}\, ,
\end{eqnarray}
where
\begin{eqnarray}
\Delta g^b_{V} &=& - 0.0001 x_h + 0.000128 x_t - 0.000025 x_h^4\, ,\nonumber\\
\Delta g^b_{A} &=& - 0.000016 x_h + 0.000128 x_t + 0.000025 x_h^4\, .\nonumber\\
\end{eqnarray}
In Eqs.~(\ref{eq:v-a-fermion}) and (\ref{eq:v-a-bquark}), we have separated out the lowest order couplings $g^f_{V,A}$, 
which are universal for all three respective fermion generations, 
\begin{eqnarray}
\label{eq:gf-V-A}
g^e_V &=& - \frac{1}{2} + 2 \sin^2\theta_W \, ,\ g^e_A = - \frac{1}{2} \, , \nonumber\\
g^u_V &=& \frac{1}{2} - \frac{4}{3}\sin^2\theta_W \, ,\ g^u_A = \frac{1}{2} \, , \nonumber\\
g^d_V &=& - \frac{1}{2} + \frac{2}{3}\sin^2\theta_W \, ,\ g^d_A = - \frac{1}{2} \, , \nonumber\\
g^{\nu}_V &=& \frac{1}{2}\, ,\ g^{\nu}_A = \frac{1}{2} \ .
\end{eqnarray}
Note that by introducing the dark photon, the couplings in Eq.~(\ref{eq:gf-V-A}) will be modified due to kinetic mixing (see Sec.~\ref{sec:C_SM}).

In principle, the weak mixing angle $\sin^2\theta_W$ should also depend on $m_{\bar Z}$, $m_{A_D}$, and $\epsilon$~\cite{Babu:1997st}. In this paper, we will take the SM value of $\sin^2\theta_W$ in the $\overline{{\rm MS}}$ scheme, $\sin^2\theta_W = 0.23122$~\cite{ParticleDataGroup:2022pth}, and check the effect of floating $\sin^2\theta_W$ when the dark photon is included.

\subsection{$Z$ boson observables}
The partial widths of the $Z$ into fermions can be expressed as~\cite{Cho:2011rk}
\begin{equation}
\label{eq:GammaZf}
\Gamma_{Z\rightarrow f{\bar f}} =
\frac{G_F m_Z^3}{6\sqrt 2 \pi}
\Big[ \Big( v_f^2+\delta_{im\,\kappa}^f \Big) C_{fV} + a_f^2 C_{fA} \Big]\Big( 1+\frac{3}{4}Q_f^2\frac{\alpha(m_Z)}{\pi} \Big)  +\Delta_{EW/QCD}^f\,,
\end{equation}
where $Q_f$ is the electric charge of fermion $f$. $C_{fV}$ and $C_{fA}$ denote corrections to the color factor in the vector
and axial-vector currents, respectively. $\Delta_{EW/QCD}^f$ are the mixed QED and QCD
corrections, and $\delta_{im\,\kappa}^f$ is the correction from the imaginary part of the loop-induced mixing of the photon and the Z boson. 
The values of these parameters are taken from Ref.~\cite{Cho:2011rk}. 

The remaining electroweak observables can be expressed in terms of these partial widths,
\begin{eqnarray}
    \sigma^0_{\rm had} &=& \frac{12\pi}{M^2_Z} \frac{\Gamma_e \Gamma_{\rm had}}{\Gamma_Z^2} \, ,\ 
    R^0_l = \frac{\Gamma_{\rm had}}{\Gamma_{l}}\, ,\ R^0_q = \frac{\Gamma_q}{\Gamma_{\rm had}}\, ,
\end{eqnarray}
where $l=e, \mu$ or $\tau$, and $q= c,b$. $\Gamma_{\rm had} = \Gamma_u + \Gamma_d + \Gamma_c + \Gamma_s + \Gamma_b$, 
and the total width $\Gamma_Z = \Gamma_{\rm had} + \Gamma_e + \Gamma_\mu + \Gamma_\tau + 3 \Gamma_{\nu}$.

The left-right asymmetry parameters $A_{\ell}$, $A_c$, and $A_b$ can be written at the tree level as
\begin{eqnarray}
A_f^{\rm tree} = \frac{2 v_f/a_f}{1 + (v_f/a_f)^2}\, .
\end{eqnarray}
Here we do not relate $A_f$ to the effective weak mixing angle $\sin^2\theta_{\rm eff}$, because the dark photon corrections to the couplings in Eq.~(\ref{eq:gf-V-A}) cannot be simply represented by a change in the Weinberg angle~\cite{Thomas:2022qhj}.

Finally, the forward-backward asymmetries, $A_{FB}^{\ell}$, $A_{FB}^{c,0}$, and $A_{FB}^{b,0}$ are given by
\begin{eqnarray}
A_{FB}^{f,0} = \frac{3}{4} A_{\ell} A_{f}\, .
\end{eqnarray}

\subsection{$W$ boson observables}
For the $W$ boson mass, we use the approximate parameterised form given in Ref.~\cite{Awramik:2003rn},
\begin{eqnarray}
\label{eq:mw-appro}
m_W &=& m_W^0 - c_1 dH - c_2 dH^2 + c_3 dH^4 + c_4 (dh - 1)  \nonumber\\
&& - c_5 d\alpha
+  c_6 dt - c_7 dt^2 - c_8 dH dt  \nonumber\\
&& + c_9 dh dt - c_{10} d\alpha_s + c_{11} d Z\, ,
\end{eqnarray}
where
\begin{eqnarray}
dH &=& \ln\left(\frac{m_h}{100\ {\rm GeV}} \right)\, , \ \ dh = \left( \frac{m_h}{100\ {\rm GeV}} \right)^2\, ,\nonumber\\ 
dt &=& \left( \frac{m_t}{174.3\ {\rm GeV}} \right)^2 - 1\, ,\ \ d\alpha = \frac{\Delta \alpha}{0.05907} - 1\, ,\nonumber \\ 
d\alpha_s &=& \frac{\alpha_s(m_Z)}{0.119} - 1\, ,\ \ d Z = \frac{m_{\bar Z}}{91.1875\ {\rm GeV}} - 1\, .
\end{eqnarray}
Here, $\Delta \alpha = \Delta\alpha_{\rm lept} + \Delta\alpha^{(5)}_{\rm had}$, where $ \Delta\alpha_{\rm lept} = 0.031497$~\cite{Steinhauser:1998rq}.
The coefficients in Eq.~(\ref{eq:mw-appro}) are given (in GeV) by:
\begin{eqnarray}
m_W^0 &=& 80.380\, ,\ \ c_1 = 0.05253\, \ \ c_2 = 0.010345\, ,\nonumber\\
c_3 &=& 0.001021\, ,\ \ c_4 = - 0.00007\, \ \ c_5 = 1.077\, ,\nonumber\\
c_6 &=& 0.5270\, ,\ \ c_7 = 0.0698\, ,\ \ c_8 =  0.004055\, ,\nonumber \\ 
c_9 &=& 0.000110\, ,\ \ c_{10} = 0.0716\, ,\ \ c_{11} = 115.0\, .
\end{eqnarray}
The total decay width of the W-boson can be parameterised by~\cite{Cho:2011rk}
\begin{equation}
\Gamma_W = 0.339 m^3_W G_F \Big[ 1 + 8.478\times 10^{-3} (2.1940 + \Delta R_W) + 0.00065 x_s \Big]\, , 
\end{equation}
where 
\begin{eqnarray}
\Delta R_W &=& -0.16 \Big\{ \ln [ 1 + ( \frac{23}{m_h (\rm GeV)} )^2 ] - \ln [ 1 + ( \frac{23}{100} )^2 ] \Big\}\, .
\end{eqnarray}

\subsection{Standard Model fit}
We follow the procedure performed in Ref.~\cite{Curtin:2014cca} and choose the free parameters
\begin{equation}
\label{eq:parameters}
    m_h, m_{\bar Z}, m_t, \alpha_s, \Delta\alpha^{(5)}_{\rm had}\, .
\end{equation}
For the Standard Model fit, $m_{\bar Z}$ is taken to be the physical mass of the $Z$ boson, $m_Z$.
These parameters are varied around their measured values to minimize the $\chi_{\rm SM}^2$, defined by
\begin{equation}
    \chi^2_{\rm SM} = V_{SM} \cdot cov^{-1} \cdot V_{SM}\, ,\ \ cov = \Sigma_{\rm exp} \cdot cor \cdot \Sigma_{\rm exp}\, .
\end{equation}
Here, $V_{\rm SM} = {\rm theory}_{\rm SM} (m_h,m_Z,m_t,\alpha_s,\Delta\alpha^{(5)}_{\rm had}) - {\rm exp}$ is the difference vector between the SM predictions and the experimental values. We use the latest experimental data~\cite{ParticleDataGroup:2022pth} as summarised in Tab.~\ref{tab:SM-fit}, while the $Z$ pole observables are taken from Ref.~\cite{Janot:2019oyi} with improved Bhabha cross section. $m_W^{\rm PDG}$ refers to the world averaged value of the $W$ boson mass from measurements at LEP, SLC, Tevatron and LHC~\footnote{We note that the ATLAS collaboration recently released an updated measurement of the $W$ boson mass, which is more precise than their previous measurement whilst remaining compatible with it~\cite{ATLAS:2023fsi}. Updating the average of this and other $m_W$ measurements is considered beyond the scope of this work, and we do not expect it to have a large effect on our conclusions.}.
$\Sigma_{\rm exp}$ is the diagonal matrix containing the experimental errors of the corresponding observables. 
The matrices $cor$ characterise the correlations among these electroweak observables, which are given by Tab.~\ref{tab:correlation-Z} and Tab.~\ref{tab:correlation-quark} for the $Z$ pole~\cite{Janot:2019oyi} and quark observables~\cite{ALEPH:2005ab}, respectively. There is also a correlation coefficient of $-0.174$ between the mass and width of the $W$ boson.

The Standard Model best fit results are given in Tab.~\ref{tab:SM-fit}, with the minimized $\chi^2$ being $\chi^2_{\rm SM} = 12.9$. If we replace the world averaged $m_W$ value by the latest CDF result~\cite{CDF:2022hxs}, the minimum $\chi^2$ soars to 68.2.
Note that the experimental value of $\alpha_s(m^2_Z)$ is not included in the fits.

\begin{table*}[!htpb]
 \renewcommand\arraystretch{1.4}
\begin{center}
\begin{tabular}{cccc} \hline \hline
            Observable                   &                 Measurement             &          Fit result ($m^{\rm PDG}_W$)      &    Fit result ($m^{\rm CDF}_W$)     \\
  \hline
         $ m_Z$ (GeV)                  &         $91.1875\pm 0.0021$       &   91.1880 $\pm$ 0.0020      &   91.1910 $\pm$ 0.0020    \\ 
         $m_h$ (GeV)                   &                125.25 $\pm$ 0.17      &    125.25 $\pm$ 0.17           &    125.24 $\pm$ 0.17     \\
         $m_t$ (GeV)                    &              172.69 $\pm$ 0.30       &   172.75 $\pm$ 0.30           &   173.09 $\pm$ 0.29     \\
 $\alpha_s(m_Z^2)$                   &  0.1179 $\pm$ 0.0009  &    0.1203 $\pm$ 0.0026     &    0.1176 $\pm$ 0.0026     \\
$\Delta\alpha^{(5)}_{\rm had}$  &         $0.02757 \pm 0.00010$         &     0.02755 $\pm$ 0.00010      &     0.02745 $\pm$ 0.00010    \\ \hline
    $\Gamma_Z$ (GeV)             &         2.4955 $\pm$ 0.0023          &          2.4963                       &     2.4953          \\
$\sigma_{\rm had}^0$ (nb)       &            41.4802 $\pm$ 0.0325           &         41.4704                    &      41.4814       \\
         $A_\ell$                           &                $0.1499\pm 0.0018$        &         0.1471                     &       0.1476             \\
         $A_b$                             &              $0.923\pm 0.020$              &        0.935                        &      0.935                \\
          $A_c$                             &                  $0.670\pm 0.027$         &         0.668                        &     0.668                \\
         $R_\ell^0$                      &                 20.7666 $\pm$ 0.0247         &        20.7529                      &      20.7358             \\
           $R_b^0$                       &             $0.21629 \pm 0.00066 $     &         0.21581                    &      0.21581          \\
           $R_c^0$                      &                 $0.1721 \pm 0.0030$       &        0.1723                        &       0.1722            \\
   $A_{FB}^{\ell,0}$                 &                 $0.0171\pm 0.0010$        &         0.0162                      &       0.0164          \\
   $A_{FB}^{b,0}$                   &                 $0.0992\pm 0.0016$         &        0.1031                     &       0.1034        \\              
     $A_{FB}^{c,0}$                 &                  $0.0707 \pm 0.0035$       &        0.0737                      &      0.0739          \\
  $m^{\rm PDG}_W$ (GeV)            &                80.377 $\pm$ 0.012      &        80.359                         &            \                    \\
    $m^{\rm CDF}_W$ (GeV)         &                $80.4335 \pm 0.0094$     &               \                             &   80.3686   \\
  $\Gamma_W$ (GeV)           &                $2.085 \pm 0.042$           &        2.091                          &    2.091      \\ \hline
  $\chi^2_{d.o.f}$                    &                                    \                     &     $12.92/(17-5)$    &    $68.23/(17-5)$     \\ \hline \hline
\end{tabular}
\caption{SM fit results.
The experimental value of $\alpha_s(m^2_Z)$ is not included in the fit.
$m_W^{\rm PDG}$ is the world averaged $W$ boson mass from measurements at LEP, SLC, Tevatron and LHC~\cite{ParticleDataGroup:2022pth}, while $m^{\rm CDF}_W$ is the recent result from CDF Collaboration~\cite{CDF:2022hxs}.}
\label{tab:SM-fit}
\end{center}
\end{table*}

\begin{table}[!htpb]
\renewcommand\arraystretch{1.3}
\begin{center} 
 \begin{tabular}{cccccc} 
                       \hline \hline  \         &     $m_Z$   &   $\Gamma_Z$   &   $\sigma^0_{\rm had}$    &    $R^0_\ell$  &   $A^{0,\ell}_{\rm FB}$    \\ \hline
                    $m_Z$    &     1.000     &           \                &                 \                        &           \           &                   \                 \\
            $\Gamma_Z$ &    $-0.0228$   &        1.000           &                 \                        &           \            &                 \                \\
$\sigma^0_{\rm had}$ &   $-0.0521$     &       $-0.3248$           &                1.000                 &          \            &                  \               \\
                 $R^0_\ell$  &   0.0332     &        0.0037           &                 0.1960                &         1.000      &                \                \\
 $A^{0,\ell}_{\rm FB}$ &  0.0549      &        0.0033           &                 0.0069               &       $-0.056$      &             1.000            \\  \hline \hline            
 \end{tabular}
 \caption{Correlations for the Z observables~\cite{Janot:2019oyi}.}
 \label{tab:correlation-Z}
\end{center}
 \end{table}

 \begin{table}[!h]
\renewcommand\arraystretch{1.3}
\begin{center}
 \begin{tabular}{ccccccc} 
              \hline \hline   \              &     $R^0_b$   &   $R^0_c$   &   $A^{0,b}_{\rm FB}$    &    $A^{0,c}_{\rm FB}$  &          $A_{b}$    &      $A_c$    \\ \hline
$R^0_b$                  &        1.00       &           \        &                 \                    &                  \                 &                \          &          \            \\
$R^0_c$                  &       $-0.18$       &        1.00     &                 \                    &                  \                 &                \           &         \            \\
$A^{0,b}_{\rm FB}$ &       $-0.10$       &         0.04    &             1.00                  &                   \                &                \           &         \              \\
$A^{0,c}_{\rm FB}$  &       0.07       &        $- 0.06$    &             0.15                 &                 1.00            &                 \          &          \             \\
$A_{b}$                   &      $-0.08$       &          0.04     &             0.06                 &               $- 0.02$            &             1.00        &         \              \\ 
$A_{c}$                   &       0.04        &         $-0.06$    &             0.01                 &                  0.04           &              0.11        &        1.00        \\  \hline \hline            
 \end{tabular}
  \caption{Correlations for the quark observables~\cite{ALEPH:2005ab}.}
  \label{tab:correlation-quark}
\end{center}
 \end{table}

\section{Constraints on the dark parameters}
\label{sec:constraints}

\subsection{Constraints on the dark photon}
\label{subsec:without chi}
\begin{figure*}[!t]
\begin{center}
\includegraphics[width=0.95\textwidth]{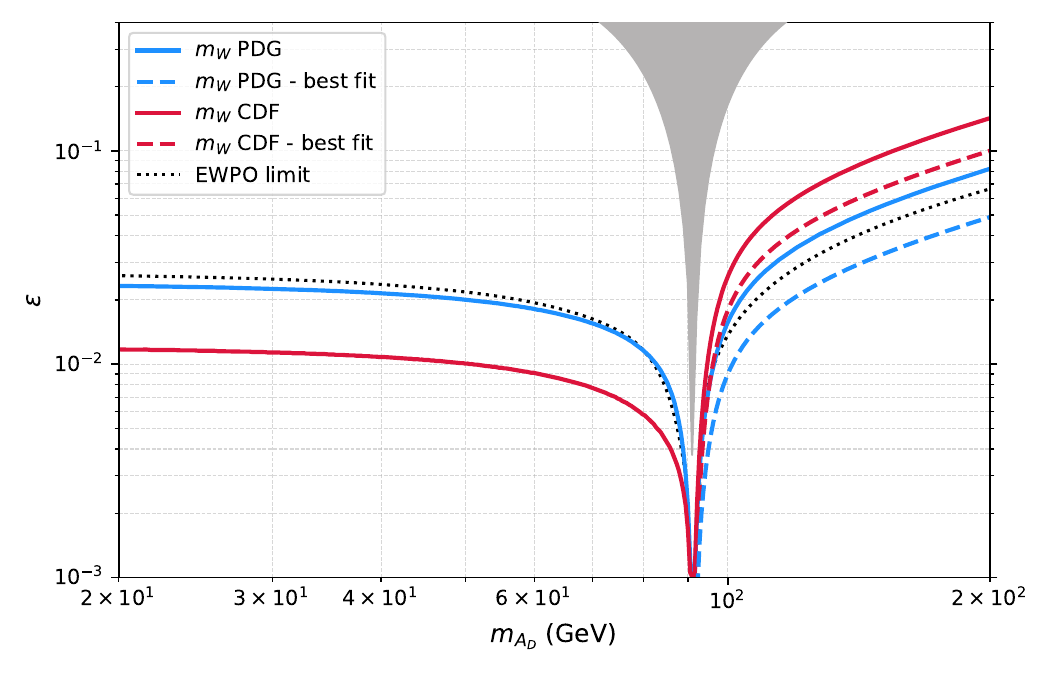}
\vspace*{-0.1cm}
\caption{ The 95\% CL exclusion constraints on $\epsilon$, excluded above. The \textit{solid} blue (red) curves represent the condition $\chi^2_{A_D} (\epsilon) - \chi^2_{\rm SM} = 3.8$, for the case in which $M_{\mathrm{w}}$ is taken to be the PDG (CDF) result. The blue (red) \textit{dashed} curves represents the parameter space that provides the \textit{best fit}, i.e the minimum $\chi^2_{A_D}$value that can be obtained by floating $\epsilon$, for which $M_{\mathrm{w}}$ is taken to be the PDG (CDF) result.
The region in grey is not accessible due to the ``eigenmass repulsion" associated with the Z mass. The EWPO limit (black dotted) is taken from Ref.~\cite{Curtin:2014cca}.
The best fit curve (red dashed) is consistent with the result in Ref.~\cite{Thomas:2022gib}.}
\label{fig:EWPO-revised}
\end{center}
\end{figure*}

We start by neglecting the couplings to dark fermions by setting $g_{\chi} = 0$. For each value of $m_{A_D}$, we adjust the mixing parameter and repeat fitting to the electroweak data by allowing the parameters in Eq.~(\ref{eq:parameters}) to vary. The minimum $\chi^2$ one can reach depends on $\epsilon$, and the 95\% CL excluded region is defined such that 
\begin{equation}
\label{eq:chi2_AD}
    \chi^2_{A_D} (\epsilon) - \chi^2_{\rm SM} \ge 3.8 \, .
\end{equation}
 The resulting upper limits on $\epsilon$ are shown in Fig.~\ref{fig:EWPO-revised}. 
In the case of the PDG value of $m_W$ (without the CDF measurement), our result (blue solid curve) is qualitatively consistent with the previous determination~\cite{Curtin:2014cca}.
 We extend this work by performing the fit with $m_W$ set to the new CDF measurement.
 We notice that the constraints on $\epsilon$ will be tightened when $m_{A_D} < m_Z$, and relaxed for $m_{A_D} > m_Z$, as shown by the red solid curve in Fig.~\ref{fig:EWPO-revised}.
 We note that the calculation of 95\% exclusion limits relies on the assumption that Wilk's theorem continues to hold even in the case of the relatively poor $\chi^2$ obtained using the CDF $m_W$ value. Our results should therefore be interpreted as indicative - a more detailed MC simulation is considered beyond the scope of this paper.
 We would expect that a new world average value of $m_W$ including the CDF result should result in an exclusion curve in between.
 
 We also search for the region in the $\epsilon - m_{A_D}$ plane in which the $\chi^2$ can be potentially improved. 
 We found that, in the region of $m_{A_D} < m_Z$, the inclusion of the dark photon will always worsen the $\chi^2$ in respect to the SM fit. In Fig.~\ref{fig:EWPO-revised},
 the blue dashed and red dashed lines represent the dark photon parameters corresponding to the maximum reduction in $\chi^2$ when fitting the PDG value and the CDF value of $m_W$, respectively. 
In the former case, the $\chi^2$ is slightly improved above the Z-pole, being $\chi^2_{A_D}= 11.38$ for $(m_{A_D}, \epsilon) = (200\ {\rm GeV}, 0.0489)$. In the case with the CDF value of $m_W$, the inclusion of a heavy dark photon significantly reduces the $\chi^2$, with $\chi^2_{A_D} = 33.7$ for $(m_{A_D}, \epsilon) = (200\ {\rm GeV}, 0.1001)$. The fit results are given in Tab.~\ref{tab:DP-fit}. 
 The predicted value of $m_W$ is $80.4060\ {\rm GeV}$, reducing the discrepancy to $2.9\ \sigma$.
 
 This result is consistent with Refs.~\cite{Thomas:2022gib, Zhang:2022nnh, Zeng:2022lkk, Cheng:2022aau}, in which the dark photon was explored as a possible explanation of the anomaly in the $W$ boson mass. Of course, while this fit represents a significant improvement over that within the SM, the $\chi^2$ is still unacceptably large.

For each point of the solid lines in Fig.~\ref{fig:EWPO-revised}, we can also obtain the fitted $m_W$ and $m_{\bar Z}$ from Eq.~(\ref{eq:chi2_AD}).
 We have also checked that, by iteratively varying $\sin^2\theta_W$ according to~\cite{ParticleDataGroup:2022pth}
 \begin{eqnarray}
 \sin^2\theta_W|_{\overline{{\rm MS}}} 
 = 1.0351 \sin^2\theta_W|_{{\rm on-shell}} 
 = 1.0351 \left( 1 - \frac{m^2_W}{m^2_{\bar Z}} \right)\, ,
 \end{eqnarray}
 and repeating the fit in Eq.~(\ref{eq:chi2_AD}), $\sin^2\theta_W$ will converge to $\sin^2\theta_W|_{\overline{{\rm MS}}} =0.23117$, and
 the changes in the resulting exclusion limits on $\epsilon$ are very small.

 \begin{table*}[!htbp]
 \renewcommand\arraystretch{1.6}
\begin{center}
\begin{tabular}{ccc}
    \hline \hline 
          Observable                  &                 Measurement          &    $(m_{A_D},\epsilon)=(200\ {\rm GeV}, 0.1001 )$     \\
  \hline
         $ m_Z$ (GeV)                &           $91.1875\pm 0.0021$         &  
        $m_{Z} =91.1894$ \\ 
         $m_h$ (GeV)                 &                125.25 $\pm$ 0.17      &    125.25 $\pm$ 0.17      \\
         $m_t$ (GeV)                 &               172.69 $\pm$ 0.30      &   172.81 $\pm$ 0.30     \\
         $\alpha_s(m_Z^2)$           &             0.1179 $\pm$ 0.0009      &    0.1205 $\pm$ 0.0026    \\
$\Delta\alpha^{(5)}_{\rm had}$       &          $0.02757 \pm 0.00010$        &     0.02759 $\pm$ 0.00010   \\ \hline
    $\Gamma_Z$ (GeV)                 &         2.4955 $\pm$ 0.0023          &     2.4936          \\
$\sigma_{\rm had}^0$ (nb)            &            41.4802 $\pm$ 0.0325         &      41.4660       \\
         $A_\ell$                    &                $0.1499\pm 0.0018$     &      0.1518             \\
         $A_b$                       &              $0.923\pm 0.020$         &      0.935                \\
          $A_c$                      &              $0.670\pm 0.027$         &     0.670                \\
         $R_\ell^0$                  &              20.7666 $\pm$ 0.0247      &      20.7640             \\
           $R_b^0$                   &             $0.21629 \pm 0.00066$     &      0.21579           \\
           $R_c^0$                   &              $0.1721 \pm 0.0030$      &       0.1723            \\
   $A_{FB}^{\ell,0}$                 &              $0.0171\pm 0.0010$       &       0.0173          \\
   $A_{FB}^{b,0}$                    &              $0.0992\pm 0.0016$       &       0.1064        \\              
     $A_{FB}^{c,0}$                  &              $0.0707 \pm 0.0035$      &      0.0763          \\
            $m^{\rm CDF}_W$ (GeV)              &             $80.4335 \pm 0.0094$      &          80.4060                   \\
     $\Gamma_W$ (GeV)                &              $2.085 \pm 0.042$        &    2.095     \\ \hline
  $\chi^2_{d.o.f}$                  &                   \                    &    33.69     \\ \hline \hline
\end{tabular}
\caption{Fit results  including the dark photon.  
The W-boson mass is taken from the latest CDF measurement.
Again, the experimental value of $\alpha_s(m^2_Z)$ is not included in the fit.
}
\label{tab:DP-fit}
\end{center}
\end{table*}

\subsection{Constraints on dark fermions}
\label{subsec:with chi}
We then switch on the dark photon's coupling to dark fermions, $g_{\chi}$, which enters only into the total $Z$ boson decay width $\Gamma_Z$. It will receive extra contributions from $\chi {\bar \chi}$ final states if $m_{\chi} < m_Z / 2$. The $Z$ boson decay width is given by:
\begin{eqnarray}
\Gamma_Z = \Gamma_{\rm had} + \Gamma_e + \Gamma_\mu + \Gamma_\tau + 3 \Gamma_{\nu} + \Gamma_{\chi}
\end{eqnarray}
where $\Gamma_{\chi}$ is the $Z\rightarrow \chi {\bar \chi}$ partial width, which depends on the dark photon mixing parameter $\epsilon$ and the coupling to dark fermions $g_{\chi}$,
\begin{eqnarray}
\label{eq:Gamma_chi}
\Gamma_{\chi} = \frac{m_Z C^2_{Z,\chi\bar{\chi}}}{12\pi} \left( 1 + \frac{2 m^2_{\chi}}{m^2_Z} \right) \sqrt{1 - \frac{4 m^2_{\chi}}{m^2_Z}}\, ,
\end{eqnarray}
with $C^2_{Z,\chi\bar{\chi}}$ defined in Eq.~(\ref{eq:C-chi-chi}). 
This feature makes it possible to determine $\epsilon$ and $g_{\chi}$ separately from electroweak observables. For a given dark photon mass, we can define the 95\% CL exclusion zone on these two parameters using ~\cite{ParticleDataGroup:2022pth}
\begin{eqnarray}
\chi^2_{A_D}(\epsilon, g_{\chi}) - \chi^2_{\rm SM} \ge 5.99\, .
\end{eqnarray}

\begin{figure*}[!htpb]
\centering
\includegraphics[width=1.06\textwidth]{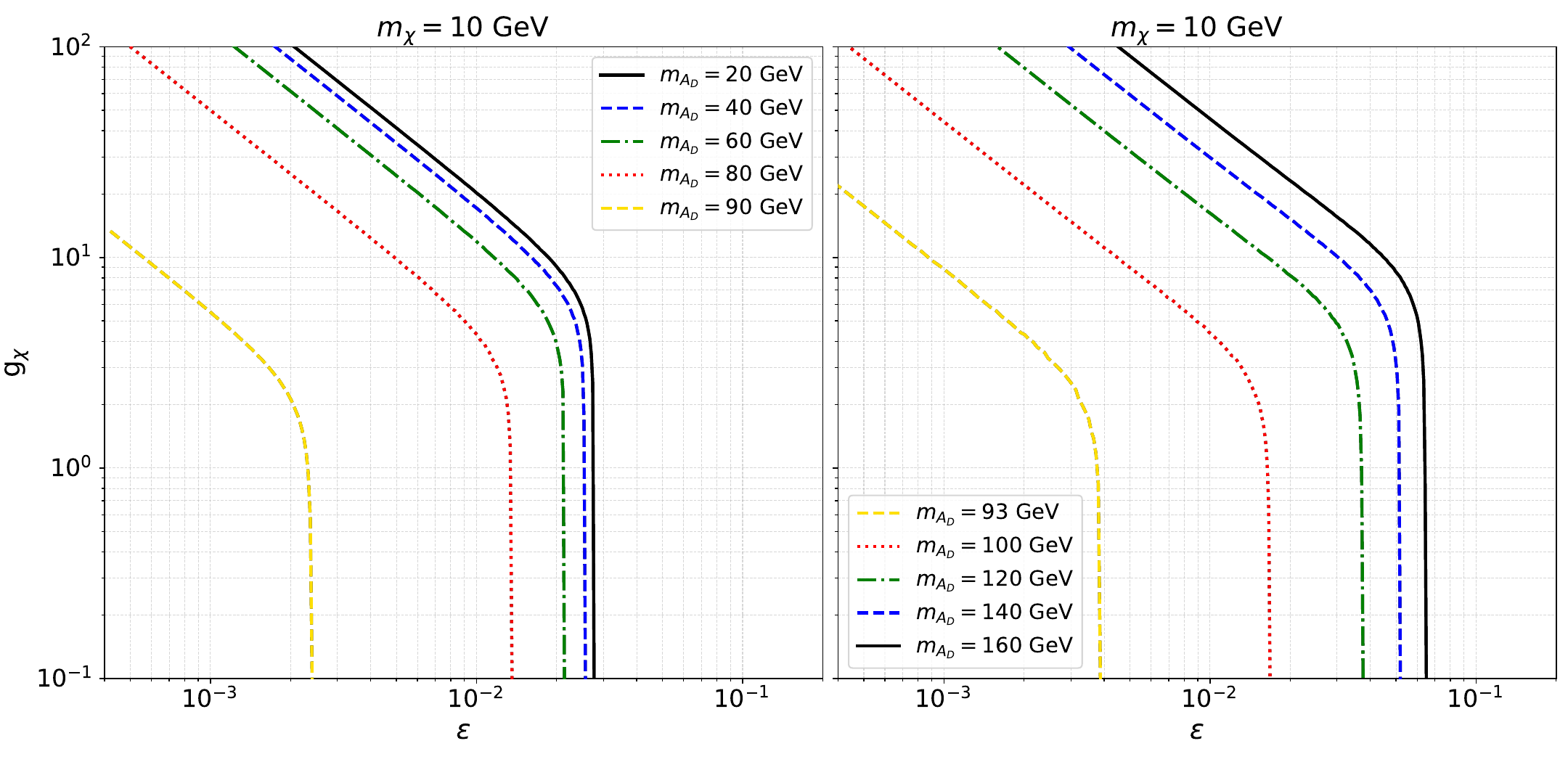}
\vspace*{-0.1cm}
\caption{ The 95\% CL exclusion constraints on dark parameters in the $g_{\chi} - \epsilon$ plane, excluded above (using $M_W^{\rm PDG}$). Each curve corresponds to $\chi^2_{A_D}(\epsilon, g_{\chi}) - \chi^2_{\rm SM} = 5.99$ for each respective $m_{A_{D}}$.}
\label{fig:g_chi_eps}
\end{figure*}

Here, we only consider the case of $m^{\rm PDG}_W$.
We first show results for heavy dark fermions with $m_{\chi}= 10\ {\rm GeV}$ in Fig.~\ref{fig:g_chi_eps} for several typical values of $m_{A_D}$.
For the region in which $m_{A_D} < m_Z$, the constraints on $g_{\chi}$ become stronger as the dark photon mass increases. Conversely, for the region in which $m_{A_D} > m_Z$, the upper bound on $g_{\chi}$ becomes relaxed.

For heavier dark fermions, the invisible partial width $\Gamma_{Z\rightarrow \chi {\bar \chi}} $ will be suppressed due to the kinematic factor in Eq.~(\ref{eq:Gamma_chi}),

\begin{equation}
K(m_{\chi}) = \left( 1 + \frac{2 m^2_{\chi}}{m^2_Z} \right) \sqrt{1 - \frac{4 m^2_{\chi}}{m^2_Z}}\, .
\end{equation}

Thus the resulting upper limits on $g_{\chi}$ shown in Fig.~\ref{fig:g_chi_eps} will be relaxed by the re-scaling factors, $R  = \sqrt{K(10\ {\rm GeV}) / K(m_{\chi})}$, as summarised in Tab.~\ref{tab:rescaling-factor}.

It is not straightforward to make comparison between the existing constraints and our results in Fig.~\ref{fig:g_chi_eps}.
The relic density, cosmic micro-wave background (CMB), and direct detection have placed constraints on the variable $y = \epsilon^2 \alpha_D (m_{\chi}/m_{A'})^4$ for light dark matter scenarios, which are insensitive to separate factors $\epsilon$ and $\alpha_D$~\cite{Izaguirre:2015yja}. 
For comparison, the EWPO~\cite{Curtin:2014cca} and the LHC limits~\cite{CMS:2019buh} on $\epsilon$ are usually multiplied by a constant $\alpha_D$ typically ranging from the QED $\alpha$ up to the perturbativity bound. 
Our results in Fig.~\ref{fig:g_chi_eps} set direct upper limits on $g_{\chi}$, which are dramatically suppressed as $\epsilon$ approaches its exclusion bound.
It would be interesting to revisit the previous constraints on $y$ by taking into account these $\epsilon$-dependent limits on $g_{\chi}$.

 \begin{table}[!htpb]
 \renewcommand\arraystretch{1.2}
\begin{center}
\begin{tabular}{ccccc}
    \hline \hline
$m_{\chi}$ (GeV)      &      10            &      20            &         30        &     40     \\  \hline
$R$               &   1.000       &     1.007       &     1.044     &  1.226 \\ \hline
\end{tabular}
\caption{Re-scaling factor to $g_{\chi}$ for heavy dark fermions.}
\label{tab:rescaling-factor}
\end{center}
\end{table}


\section{Conclusions}
\label{sec:conclusion}
In this work, we revisited the constraints on the dark photon from electroweak precision observables, by investigating the impact of the latest CDF $m_W$ measurement on the exclusion limits of the mixing parameter $\epsilon$. The upper bounds of $\epsilon$ become tightened in the region $m_{A_D} < m_Z$, and weakened for $m_{A_D} > m_Z$. We also searched for regions in the dark photon parameter space in which the inclusion of the dark photon could potentially improve the agreement between theory and experiment. 

We also introduced dark photon couplings to fermionic dark matter particles $\chi$. Due to kinetic mixing, the $Z$ boson also decays invisibly to $\chi\bar{\chi}$ in addition to the neutrino final states. Fitting to the electroweak precision observables then sets exclusion limits on $g_{\chi}$ for massive dark fermions, which are rather strigent when the dark photon mass gets close to the $Z$ boson mass. These upper bounds on $g_{\chi}$ become weaker as $m_{\chi}$ increases.

In the future, the proposed CEPC~\cite{CEPCStudyGroup:2018ghi}, FCC-ee~\cite{FCC:2018evy}, ILC~\cite{ILC:2019gyn} and CLIC~\cite{CLICdp:2018cto} are expected to measure some of the electroweak observables with significantly increased precision, which could improve the current constraints on the dark sector. Moreover, the analysis presented in this work can also be extended to pseudo-Dirac, scalar, and asymmetric dark matter scenarios.

\section*{Acknowledgments}
We thank Dipan Sengupta and Stefania Gori for helpful discussions.
This work was supported by the University of Adelaide and the Australian Research Council through the Centre of Excellence for Dark Matter Particle Physics (CE200100008). This work was also supported by an Australian Government Research Training Program Scholarship.

\section*{References}

\bibliographystyle{iopart-num}
\bibliography{bibliography}

\providecommand{\newblock}{}
\begin{thebibliography}{10}
\expandafter\ifx\csname url\endcsname\relax
  \def\url#1{{\tt #1}}\fi
\expandafter\ifx\csname urlprefix\endcsname\relax\def\urlprefix{URL }\fi
\providecommand{\eprint}[2][]{\url{#2}}

\bibitem{Fabbrichesi:2020wbt}
Fabbrichesi M, Gabrielli E and Lanfranchi G 2020  (\textit{Preprint}
  \eprint{2005.01515})

\bibitem{Filippi:2020kii}
Filippi A and De~Napoli M 2020 {\em Rev. Phys.\/} {\bf 5} 100042
  (\textit{Preprint} \eprint{2006.04640})

\bibitem{Graham:2021ggy}
Graham M, Hearty C and Williams M 2021 {\em Ann. Rev. Nucl. Part. Sci.\/} {\bf
  71} 37--58 (\textit{Preprint} \eprint{2104.10280})

\bibitem{Merkel:2014avp}
Merkel H {\em et~al.\/} 2014 {\em Phys. Rev. Lett.\/} {\bf 112} 221802
  (\textit{Preprint} \eprint{1404.5502})

\bibitem{LHCb:2019vmc}
Aaij R {\em et~al.\/} (LHCb) 2020 {\em Phys. Rev. Lett.\/} {\bf 124} 041801
  (\textit{Preprint} \eprint{1910.06926})

\bibitem{Banerjee:2019pds}
Banerjee D {\em et~al.\/} 2019 {\em Phys. Rev. Lett.\/} {\bf 123} 121801
  (\textit{Preprint} \eprint{1906.00176})

\bibitem{BaBar:2017tiz}
Lees J~P {\em et~al.\/} (BaBar) 2017 {\em Phys. Rev. Lett.\/} {\bf 119} 131804
  (\textit{Preprint} \eprint{1702.03327})

\bibitem{CMS:2019buh}
Sirunyan A~M {\em et~al.\/} 2020 {\em Phys. Rev. Lett.\/} {\bf 124} 131802
  (\textit{Preprint} \eprint{1912.04776})

\bibitem{Hunt-Smith:2023sdz}
Hunt-Smith N~T, Melnitchouk W, Sato N, Thomas A~W, Wang X~G and White M~J
  (Jefferson Lab Angular Momentum (JAM)) 2023 {\em JHEP\/} {\bf 09} 096
  (\textit{Preprint} \eprint{2302.11126})

\bibitem{Abdullahi:2023tyk}
Abdullahi A~M, Hostert M, Massaro D and Pascoli S 2023 {\em Phys. Rev. D\/}
  {\bf 108} 015032 (\textit{Preprint} \eprint{2302.05410})

\bibitem{APEX:2011dww}
Abrahamyan S {\em et~al.\/} (APEX) 2011 {\em Phys. Rev. Lett.\/} {\bf 107}
  191804 (\textit{Preprint} \eprint{1108.2750})

\bibitem{Battaglieri:2014hga}
Battaglieri M {\em et~al.\/} 2015 {\em Nucl. Instrum. Meth. A\/} {\bf 777}
  91--101 (\textit{Preprint} \eprint{1406.6115})

\bibitem{Beranek:2013yqa}
Beranek T, Merkel H and Vanderhaeghen M 2013 {\em Phys. Rev. D\/} {\bf 88}
  015032 (\textit{Preprint} \eprint{1303.2540})

\bibitem{Pospelov:2008zw}
Pospelov M 2009 {\em Phys. Rev. D\/} {\bf 80} 095002 (\textit{Preprint}
  \eprint{0811.1030})

\bibitem{Davoudiasl:2012qa}
Davoudiasl H, Lee H~S and Marciano W~J 2012 {\em Phys. Rev. Lett.\/} {\bf 109}
  031802 (\textit{Preprint} \eprint{1205.2709})

\bibitem{Hook:2010tw}
Hook A, Izaguirre E and Wacker J~G 2011 {\em Adv. High Energy Phys.\/} {\bf
  2011} 859762 (\textit{Preprint} \eprint{1006.0973})

\bibitem{Curtin:2014cca}
Curtin D, Essig R, Gori S and Shelton J 2015 {\em JHEP\/} {\bf 02} 157
  (\textit{Preprint} \eprint{1412.0018})

\bibitem{Kribs:2020vyk}
Kribs G~D, McKeen D and Raj N 2021 {\em Phys. Rev. Lett.\/} {\bf 126} 011801
  (\textit{Preprint} \eprint{2007.15655})

\bibitem{Thomas:2021lub}
Thomas A~W, Wang X~G and Williams A~G 2022 {\em Phys. Rev. D\/} {\bf 105}
  L031901 (\textit{Preprint} \eprint{2111.05664})

\bibitem{Yan:2022npz}
Yan B 2022 {\em Phys. Lett. B\/} {\bf 833} 137384 (\textit{Preprint}
  \eprint{2203.01510})

\bibitem{Thomas:2022qhj}
Thomas A~W, Wang X~G and Williams A~G 2022 {\em Phys. Rev. Lett.\/} {\bf 129}
  011807 (\textit{Preprint} \eprint{2201.06760})

\bibitem{Thomas:2022gib}
Thomas A~W and Wang X~G 2022 {\em Phys. Rev. D\/} {\bf 106} 056017
  (\textit{Preprint} \eprint{2205.01911})

\bibitem{Hosseini:2022urq}
Hosseini Y and Najafabadi M~M 2022 {\em Phys. Rev. D\/} {\bf 106} 015028
  (\textit{Preprint} \eprint{2202.10058})

\bibitem{Davoudiasl:2012ag}
Davoudiasl H, Lee H~S and Marciano W~J 2012 {\em Phys. Rev. D\/} {\bf 85}
  115019 (\textit{Preprint} \eprint{1203.2947})

\bibitem{Wang:2023css}
Wang X~G and Thomas A~W 2023 {\em J. Phys. G\/} {\bf 50} 085001
  (\textit{Preprint} \eprint{2301.08367})

\bibitem{McCullough:2022hzr}
McCullough M, Moore J and Ubiali M 2022 {\em JHEP\/} {\bf 08} 019
  (\textit{Preprint} \eprint{2203.12628})

\bibitem{CDF:2022hxs}
Aaltonen T {\em et~al.\/} 2022 {\em Science\/} {\bf 376} 170--176

\bibitem{Zhang:2022nnh}
Zhang K~Y and Feng W~Z 2023 {\em Chin. Phys. C\/} {\bf 47} 023107
  (\textit{Preprint} \eprint{2204.08067})

\bibitem{Zeng:2022lkk}
Zeng Y~P, Cai C, Su Y~H and Zhang H~H 2023 {\em Phys. Rev. D\/} {\bf 107}
  056004 (\textit{Preprint} \eprint{2204.09487})

\bibitem{Cheng:2022aau}
Cheng Y, He X~G, Huang F, Sun J and Xing Z~P 2022 {\em Phys. Rev. D\/} {\bf
  106} 055011 (\textit{Preprint} \eprint{2204.10156})

\bibitem{Lu:2022bgw}
Lu C~T, Wu L, Wu Y and Zhu B 2022 {\em Phys. Rev. D\/} {\bf 106} 035034
  (\textit{Preprint} \eprint{2204.03796})

\bibitem{deBlas:2022hdk}
de~Blas J, Pierini M, Reina L and Silvestrini L 2022 {\em Phys. Rev. Lett.\/}
  {\bf 129} 271801 (\textit{Preprint} \eprint{2204.04204})

\bibitem{Boehm:2003hm}
Boehm C and Fayet P 2004 {\em Nucl. Phys. B\/} {\bf 683} 219--263
  (\textit{Preprint} \eprint{hep-ph/0305261})

\bibitem{Hambye:2019dwd}
Hambye T, Tytgat M~H~G, Vandecasteele J and Vanderheyden L 2019 {\em Phys. Rev.
  D\/} {\bf 100} 095018 (\textit{Preprint} \eprint{1908.09864})

\bibitem{Hu:2000ke}
Hu W, Barkana R and Gruzinov A 2000 {\em Phys. Rev. Lett.\/} {\bf 85}
  1158--1161 (\textit{Preprint} \eprint{astro-ph/0003365})

\bibitem{Veltmaat:2016rxo}
Veltmaat J and Niemeyer J~C 2016 {\em Phys. Rev. D\/} {\bf 94} 123523
  (\textit{Preprint} \eprint{1608.00802})

\bibitem{Luongo:2018lgy}
Luongo O and Muccino M 2018 {\em Phys. Rev. D\/} {\bf 98} 103520
  (\textit{Preprint} \eprint{1807.00180})

\bibitem{Carney:2022gse}
Carney D {\em et~al.\/} 2023 {\em SciPost Phys. Core\/} {\bf 6} 075
  (\textit{Preprint} \eprint{2203.06508})

\bibitem{Planck:2015fie}
Ade P~A~R {\em et~al.\/} (Planck) 2016 {\em Astron. Astrophys.\/} {\bf 594} A13
  (\textit{Preprint} \eprint{1502.01589})

\bibitem{Izaguirre:2015yja}
Izaguirre E, Krnjaic G, Schuster P and Toro N 2015 {\em Phys. Rev. Lett.\/}
  {\bf 115} 251301 (\textit{Preprint} \eprint{1505.00011})

\bibitem{Fayet:1980ad}
Fayet P 1980 {\em Phys. Lett. B\/} {\bf 95} 285--289

\bibitem{Fayet:1980rr}
Fayet P 1981 {\em Nucl. Phys. B\/} {\bf 187} 184--204

\bibitem{Holdom:1985ag}
Holdom B 1986 {\em Phys. Lett. B\/} {\bf 166} 196--198

\bibitem{Okun:1982xi}
Okun L~B 1982 {\em Sov. Phys. JETP\/} {\bf 56} 502

\bibitem{Hagiwara:1994pw}
Hagiwara K, Matsumoto S, Haidt D and Kim C~S 1994 {\em Z. Phys. C\/} {\bf 64}
  559--620 [Erratum: Z.Phys.C 68, 352 (1995)] (\textit{Preprint}
  \eprint{hep-ph/9409380})

\bibitem{Cho:1999km}
Cho G~C and Hagiwara K 2000 {\em Nucl. Phys. B\/} {\bf 574} 623--674
  (\textit{Preprint} \eprint{hep-ph/9912260})

\bibitem{Cho:2011rk}
Cho G~C, Hagiwara K, Matsumoto Y and Nomura D 2011 {\em JHEP\/} {\bf 11} 068
  (\textit{Preprint} \eprint{1104.1769})

\bibitem{Babu:1997st}
Babu K~S, Kolda C~F and March-Russell J 1998 {\em Phys. Rev. D\/} {\bf 57}
  6788--6792 (\textit{Preprint} \eprint{hep-ph/9710441})

\bibitem{ParticleDataGroup:2022pth}
Workman R~L {\em et~al.\/} (Particle Data Group) 2022 {\em PTEP\/} {\bf 2022}
  083C01

\bibitem{Awramik:2003rn}
Awramik M, Czakon M, Freitas A and Weiglein G 2004 {\em Phys. Rev. D\/} {\bf
  69} 053006 (\textit{Preprint} \eprint{hep-ph/0311148})

\bibitem{Steinhauser:1998rq}
Steinhauser M 1998 {\em Phys. Lett. B\/} {\bf 429} 158--161 (\textit{Preprint}
  \eprint{hep-ph/9803313})

\bibitem{Janot:2019oyi}
Janot P and Jadach S 2020 {\em Phys. Lett. B\/} {\bf 803} 135319
  (\textit{Preprint} \eprint{1912.02067})

\bibitem{ATLAS:2023fsi}
ATLAS 2023 {Improved W boson Mass Measurement using 7 TeV Proton-Proton
  Collisions with the ATLAS Detector}
  \urlprefix\url{https://inspirehep.net/files/d8239d2613f5889910b01e6c77cdb9b3}

\bibitem{ALEPH:2005ab}
Schael S {\em et~al.\/} (ALEPH, DELPHI, L3, OPAL, SLD, LEP Electroweak Working
  Group, SLD Electroweak Group, SLD Heavy Flavour Group) 2006 {\em Phys.
  Rept.\/} {\bf 427} 257--454 (\textit{Preprint} \eprint{hep-ex/0509008})

\bibitem{CEPCStudyGroup:2018ghi}
Dong M {\em et~al.\/} (CEPC Study Group) 2018  (\textit{Preprint}
  \eprint{1811.10545})

\bibitem{FCC:2018evy}
Abada A {\em et~al.\/} (FCC) 2019 {\em Eur. Phys. J. ST\/} {\bf 228} 261--623

\bibitem{ILC:2019gyn}
Aihara H {\em et~al.\/} (ILC) 2019  (\textit{Preprint} \eprint{1901.09829})

\bibitem{CLICdp:2018cto}
Charles T~K {\em et~al.\/} (CLICdp, CLIC) 2018  {\bf 2/2018} (\textit{Preprint}
  \eprint{1812.06018})

\end{thebibliography}

\end{document}